\pdfoutput=1
\documentclass[prl,twocolumn,notitlepage,superscriptaddress,amsmath,amssymb,floatfix,longbibliography]{revtex4-2}
\usepackage[T1]{fontenc}
\usepackage{float}
\usepackage{color}
\usepackage{amsmath}
\usepackage{amssymb}
\usepackage{bm}
\usepackage{graphicx}
\usepackage[usenames,dvipsnames]{xcolor}
\usepackage[colorlinks=true,urlcolor=blue,linkcolor=blue,citecolor=blue]{hyperref}
\usepackage{array}
\definecolor{bleu}{rgb}{0.16,0.2.5,0.36}

\newcommand{\rs}[1]{\textcolor{black}{#1}}

\usepackage{lipsum} 
\usepackage{amssymb}
\usepackage{booktabs}
\usepackage{comment}

\begin{document}

\title{Combinatorial Design of Floppy Modes and Frustrated Loops in Metamaterials}

\author{Wenfeng Liu}
\affiliation{Institute of Physics, Universiteit van Amsterdam, Amsterdam 1098 XH, The Netherlands}

\author{Tomer A. Sigalov}
\affiliation{School of Mechanical Engineering, Tel Aviv University, Tel Aviv 69978, Israel}

\author{Corentin Coulais}
\affiliation{Institute of Physics, Universiteit van Amsterdam, Amsterdam 1098 XH, The Netherlands}

\author{Yair Shokef}
\email{shokef@tau.ac.il}
\affiliation{School of Mechanical Engineering, Tel Aviv University, Tel Aviv 69978, Israel}
\affiliation{School of Physics and Astronomy, Tel Aviv University, Tel Aviv 69978, Israel}
\affiliation{Center for Computational Molecular and Materials Science, Tel Aviv University, Tel Aviv 69978, Israel}
\affiliation{Center for Physics and Chemistry of Living Systems, Tel Aviv University, 69978, Tel Aviv, Israel}
\affiliation{International Institute for Sustainability with Knotted Chiral Meta Matter (WPI-SKCM$^2$), Hiroshima University, Higashi-Hiroshima, Hiroshima 739-8526, Japan}

\begin{abstract}
Metamaterials are a promising platform for a range of applications, from shock absorption to mechanical computing. These functionalities typically rely on floppy modes or mechanically frustrated loops, both of which are difficult to design. \rs{In particular, how to design multiple modes or loops with target deformations remains an open problem.} We introduce a combinatorial approach that allows \rs{us} to create an arbitrarily large number of floppy modes and frustrated loops. The design freedom of the mode shapes enables us to easily introduce kinematic incompatibility to turn them into frustrated loops. We demonstrate that floppy modes can be sequentially buckled by using a specific instance of elastoplastic buckling. \rs{We} utilize our combinatorial floppy chains and frustrated loops to achieve matrix-vector multiplication in materia. Our findings bring about new principles for the design and the use of floppiness and geometric frustration in soft matter and metamaterials.
\end{abstract}

\maketitle

\emph{Introduction --} 
In soft matter, mechanical properties are often rooted in floppy modes and geometrically frustrated states of self-stress. Examples range from protein allostery~\cite{rocks2017}, granular packings~\cite{vanhecke2009, liu2010}  and colloidal glasses~\cite{chen2010, ghosh2010} to amorphous solids~\cite{alexander1998, wyart2005, vogel2025}, biopolymer networks~\cite{broedersz2011, broedersz2014, zhou2018} and mechanical metamaterials~\cite{kane2014a, lubensky2015phonons, Tang2024}. Mechanical metamaterials are a platform of choice to study floppy and frustrated modes~\cite{bertoldi_flexible_review_2017}. Once one understands those modes, one can in turn harness them to achieve on-demand unusual properties~\cite{reis2015a}. Prime examples are topological wave guiding and stress focusing~\cite{Paulose_PNAS2015,  huber2016topological, rocklin2016, sirota2020, jin2020, Bilal_AdvMater2021, Widstrand_EML2024}, vibration and shock absorption~\cite{shan2015multistable, dykstra2023buckling, liu2024harnessing}, \rs{control of fracture~\cite{widstrand2023stress, Wang_preprint2024, de2025architecting}}, shape morphing~\cite{coulais2016metacube, dieleman_jigsaw_2020, shapemorphingreview2025, Melio2025} and mechanical computing~\cite{treml2018origami, yasuda2021mechanical, 
berry2022, kwakernaak2023, louvet_matrix_product, alu2025}. 

Floppy modes are deformations that cost negligible elastic energy~\cite{haghpanah2016, Deng_domain_walls_2020, bossart_oligomodal_2021, czajkowski_conformal_2022, van_mastrigt_machine_2022, meng2022, van_mastrigt_emergent_2023, sirote_compatible}. Frustrated stressed states are often configured in loops~\cite{kang2014complex, meeussen2020supertriangles, meeussen_NJP_2020, pisanty2021, guo2023non, chaco, sirote_defects}, where these low-energy deformations are geometrically impossible. Both floppy modes and frustrated loops can be used to channel deformations and stresses efficiently. \rs{Yet,} designing metamaterials with multiple, precisely controlled modes or loops remains a challenge. \rs{Existing works rely on topological~\cite{kane2014a, lubensky2015phonons, Paulose_PNAS2015, zhou2018, zhou2019topological, Tang2024} or computational~\cite{rocks2017, van_mastrigt_machine_2022, van_mastrigt_emergent_2023} approaches and exhibit limited design freedom on the number of modes and loops, and on their spatial structure}.

\rs{Here,} we open up the design space of combinatorial metamaterials, by introducing a theoretical \rs{strategy} that allows \rs{unpredecented} freedom in setting the number and shapes of floppy modes and frustrated loops. We leverage this approach to create metamaterials that exhibit advanced sequential response upon uniaxial compression and mechanical computing in the form of matrix-vector multiplication. Thus, our work bridges abstract principles and practical applications in soft programmable materials and adaptive architectures. 

\begin{figure}[t!]
\centering
\includegraphics[width=\columnwidth]{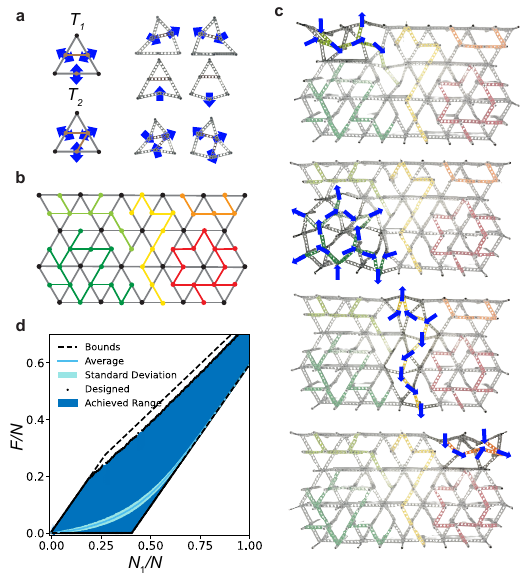}
\caption{\textbf{Floppy modes in metamaterials with perfect hinges and rigid bonds.} \textbf{a,} Triangular blocks with spins (\rs{blue arrows}) describing the direction of displacement of edge nodes, and bonds (brown) constraining spins to move in alternating manner. One internal bond in Block $T_1$ constrains two spins, thus the third is independent, resulting in two floppy modes. Block $T_2$ has two internal bonds, thus all spins displace together in one floppy mode. \textbf{b,} Metamaterial made of $T_1$ and $T_2$ blocks. Chains of connected spins constraining each other's motion are individually colored; open chains (light green, yellow, orange) represent nodes moving together in a floppy mode, and so do chains that contain closed loops of even length (dark green). Chains with odd loop (red) are mechanically frustrated and hence rigid. \textbf{c,} Experimental demonstration of the four floppy modes of the system in b. \textbf{d,} Normalized number $F/N$ of floppy modes vs. normalized number $N_1/N$ of $T_1$ blocks for a lattice of $N=210$ blocks. The values of $F$ span almost the entire range between the lower and upper theoretical bounds. The average number of floppy modes in randomly generated systems exhibits small fluctuations and is closer to the lower bound.}
\label{fig:1}
\end{figure}

\emph{Building Blocks and Spin Model --}
We consider two-dimensional mechanical metamaterials composed of triangular building blocks, which are constructed from rigid bonds connected at freely rotating hinges. Each block consists of three corner nodes and three edge nodes. The nodes are connected by six rigid perimeter bonds. In addition, $T_1$ triangles have one internal bond connecting two adjacent edge nodes, while $T_2$ triangles have two such internal bonds (Fig.~\ref{fig:1}a). For floppy modes, in the small deformation limit, the corner nodes remain stationary while edge nodes move perpendicular to their respective edges. This key insight allows us to introduce a spin model to describe the system's deformations. In this model, we assign a \rs{binary} spin-like variable to each edge node, describing whether it displaces into the triangle or out of it. Effective antiferromagnetic interactions between spins are introduced by internal bonds, since they constrain connected edge nodes to move in alternating directions with respect to the triangle. \rs{Thus, a floppy, or zero-energy deformation corresponds to a spin configuration which satisfies all the antiferromagnetic interactions, while frustration appears whenever there is a closed loop with an odd number of bonds.}

\emph{Design of Floppy Modes --}
Using our spin model, we can design metamaterials with varying numbers and shapes of floppy modes. By connecting spins, we create \emph{chains} of nodes (Fig.~\ref{fig:1}b \rs{and Video~1}) that can move together, independently from other chains in the system (Fig.~\ref{fig:1}c). The mutual direction of motion of the nodes is determined by the internal bonds connecting them along the chain. This demonstrates our ability to create chains with complex, predefined shapes, and deformations. Furthermore, an inherent property caused due to the antiferromagnetic interaction between spins is that a closed loop with an even number of such bonds allows the corresponding spins to move together (Fig.~\ref{fig:1}b,c, dark green chain), while a loop with an odd length renders the chain frustrated and rigidifies the connected spins (Fig.~\ref{fig:1}b, red chain). 

To validate our approach and the spin model, we explicitly write the rigidity matrix for the actual geometries that we consider, we identify from the matrix how many floppy modes the system has, and then verify that the theoretically-predicted floppy modes from the model indeed give zero when multiplied by the rigidity matrix. We construct a physical model using LEGO\textsuperscript{\tiny\textregistered} beams and axles, which closely approximates our theoretical system of flexible hinges and rigid bonds. We successfully actuate the floppy modes by displacing nodes, which validates our design methodology and shows that these modes are, in fact, independent zero-energy deformation modes (Fig.~\ref{fig:1}c and Video~1). 

\emph{Number of Modes --}
On top of designing the spatial form of floppy modes, the spin model allows to quantify their number in such metamaterials; the number of floppy modes is the number of independent degrees of freedom in the system. Each spin is a degree of freedom, and each bond is a constraint that reduces the number of independent degrees of freedom. However, a bond which closes a loop is redundant and does not add a constraint, while chains with an odd loop are rigid and do not possess a degree of freedom. A lattice of $N$ triangles and of perimeter $P$ contains $\frac{3N}{2}+\frac{P}{2}$ edge nodes, or spins. Denoting the numbers of $T_1$ and $T_2$ triangles by $N_1$ and $N_2$, respectively, the number of internal bonds is $N_1+2N_2$. Using $N=N_1+N_2$, \rs{this analysis yields} an exact expression for the number of floppy modes, $F = N_1 - \frac{N}{2} + \frac{P}{2} + L - R$, where $L$ is the number of closed loops of internal bonds, and $R$ is the number of rigid chains.

For given lattice size and shape that set $N$ and $P$, and for given values $N_1$ and $N_2$ of $T_1$ and $T_2$ blocks, the number of floppy modes can take a wide range (Fig.~\ref{fig:1}d), depending on how the blocks are arranged to form chains and loops, thus changing the value of $L-R$. The metamaterial design can lead to any number of modes between the bounds \rs{(see Appendix)},
\begin{align}
F \ge \begin{cases}
0, & N_1 \le \frac{N-P}{2} \nonumber \\
N_1-\frac{N-P}{2}, & N_1 \ge \frac{N-P}{2}
\end{cases} ,
\\
F \le \begin{cases}
N_1+1, & N_1 \le \frac{3P}{2}-3\\
\frac{2N_1}{3} + \frac{P}{2}, & N_1 \ge \frac{3P}{2}-3
\end{cases} .
\label{eq:bounds}
\end{align}
If most triangles are $T_2$ $\left( {\rm namely,} \ N_1 \lesssim \frac{N}{2} \right)$, the triangles can be arranged such that the metamaterial will be rigid, without any floppy modes, $F=0$. For any ratio of $T_1$ and $T_2$ triangles, there are always designs that give an extensive number of floppy modes $F \propto N$, and if most blocks are $T_1$ $\left( N_1 \gtrsim \frac{N}{2} \right)$, this number is necessarily extensive. Finally, we observe that by randomly positioning and orienting the blocks, the number of floppy modes is closer to the lower bound (Fig.~\ref{fig:1}d\rs{, and see Appendix}). \rs{In summary, our combinatorial strategy allows us to design an arbitrary number of modes in a systematic way and with complex target shapes, unlike existing approaches thus far~\cite{kane2014a, Paulose_PNAS2015, lubensky2015phonons, rocks2017, zhou2018, zhou2019topological, van_mastrigt_machine_2022, van_mastrigt_emergent_2023, Tang2024}.}

\begin{figure}[t]
\centering
\includegraphics[width=\columnwidth]{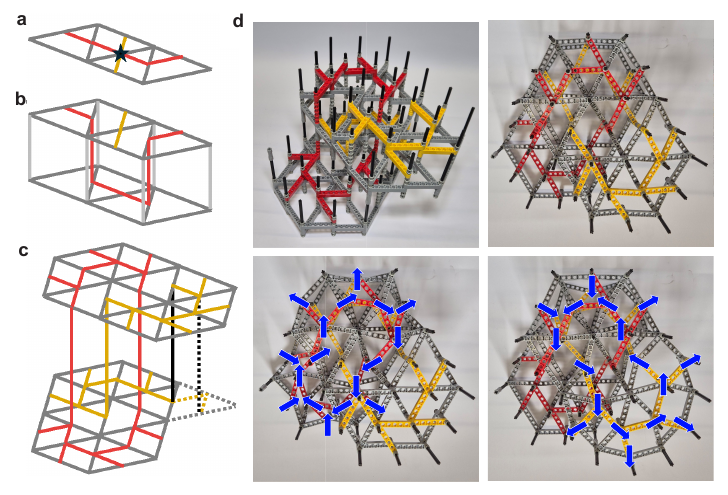}
\caption{\textbf{Linked floppy modes and three-dimensional rigid chains.} \textbf{a,} In the plane, two crossing chains (red, yellow) share an edge node (black star) that couples them. \textbf{b,}~Connecting parallel layers allows chains to bypass each other. \textbf{c,} Three-dimensional metamaterials enable knotted topologies, like catenated floppy chains, without contact between them. For clarity, the vertical connectors at all corner nodes are not shown. Moving the solid black vertical connector to its dotted position rigidifies the yellow loop without changing the intra-layer structure. \textbf{d,} LEGO\textsuperscript{\tiny\textregistered} realization of linked loops and their individual actuation.}
\label{fig:lego_3D}
\end{figure}

\emph{Mode Crossing --}
Although we can design the shapes of multiple floppy modes, these modes cannot cross each other. For that, they would have to share a node at the crossing point. That node's motion would cause the simultaneous actuation of the two modes, and they would thus, in fact, be a single mode (Fig.~\ref{fig:lego_3D}a). To allow modes to cross, we generalize our approach to a three-dimensional layered system where modes can bypass each other via the third dimension (Fig.~\ref{fig:lego_3D}b). We couple adjacent layers with vertical connectors and allow only planar displacements within each layer. By connecting corner nodes across layers, we remove shearing and twisting. 

Connecting specific edge nodes, we can design floppy modes of arbitrary topology, like catenated floppy loops (Fig.~\ref{fig:lego_3D}c) that may be individually actuated since they do not touch each other (Fig.~\ref{fig:lego_3D}d and Video~2). Furthermore, chains that in individual layers are floppy may now form rigid loops that traverse multiple layers in a manner that is sensitive to the positions of the vertical connectors. For instance, moving the black connector in Fig.~\ref{fig:lego_3D}c to its dotted position changes the parity of the yellow loop from even and thus floppy to odd and thus rigid, without changing the structure of any of the layers. \rs{Generalizing the analysis of the number of floppy modes to such layered metamaterials is straightforward (see Appendix).}

By extending our design methodology to layered lattices, we significantly expand the range of achievable functionalities and topologies in our mechanical metamaterials and (literally) bridge the gap between planar designs and fully three-dimensional structures. 

In the remainder of the paper, we utilize floppy modes \rs{and frustrated loops} to achieve functional responses such as (i) the sequential buckling of designer-shaped floppy modes \rs{under uniform compression} and (ii) matrix-vector multiplication \rs{with multiple inputs and outputs under local actuation}. \rs{In both, the enhanced design freedom of our combinatorial approach leads to more advanced responses than in earlier works.}

\emph{Floppy Mode Buckling --} 
To go beyond \emph{local} actuation of individual modes, we exploit the buckling instability as a mechanism to actuate floppy modes by uniform \emph{global} compression. Unlike existing metamaterials based on buckling, with one mode~\cite{bertoldi2010negative}, a large number of modes along straight lines~\cite{lubensky2015phonons, bossart_oligomodal_2021, van_mastrigt_machine_2022, liu2024harnessing, liu2025tuning}, or limited modes in a hierarchical design~\cite{coulais2018multi, wu2024zero}, our modes can be shaped into tortuous chains or loops \rs{based on the combinatorial design of the metamaterial}, and their buckling onset is determined by the length of the chains and by the parity of the loops (Video~3). 

\begin{figure}[t]
\centering
\includegraphics[width=\columnwidth]{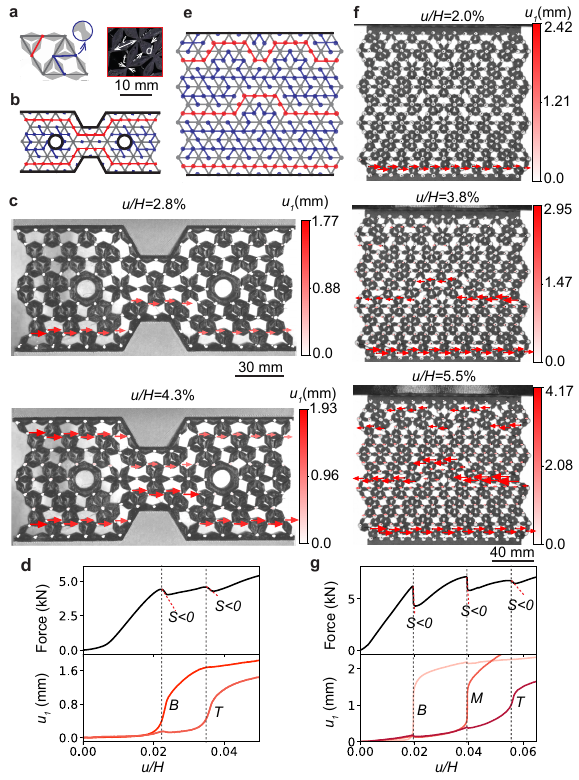}
\caption{\textbf{Sequential actuation of floppy modes.} \textbf{a,} Theoretical model, corresponding metamaterial geometry \rs{and 3D printed components}, with rigid bonds replaced by diamonds. \rs{\textbf{b,} Irregularly shaped metamaterial with two identical floppy chains (red), separated by disconnected rigid chains (blue).} \textbf{e,}~Metamaterial with three floppy chains (red) of varying length, separated by disconnected rigid chains (blue). \textbf{c,~f,}~Corresponding plastic metamaterials at different stages of compression. \rs{Arrows of red intensity} indicate horizontal displacement. \textbf{d, g,} Corresponding vertical force and average horizontal displacement, $u_1$ of each floppy chain vs. global compressing stroke $u/H$. $B$, $M$, and $T$ denote the bottom, middle, and top floppy chains. The sign of the slope ($S$) of force vs compression is detected (dashed lines).}
\label{fig:sequential_YB}
\end{figure}

We demonstrate floppy mode actuation in a 3D-printed metamaterial made of diamond-shaped beams (Fig.~\ref{fig:sequential_YB}a). The thin connectors at the diamond ends allow a relatively flexible rotation compared to the substantially greater resistance to deforming the individual diamonds. \rs{We first demonstrate two identical floppy chains separated by rigid regions in an irregular-shaped metamaterial with two isolated holes inside (Fig.~\ref{fig:sequential_YB}b)}

\emph{Sequential Yield Buckling --} 
We separate the excitation of the \rs{identical} modes by exploiting yield buckling~\cite{liu2024harnessing}, which is rooted in the symbiotic occurrence of plastic yielding and buckling. To combine plasticity with the floppy modes, \rs{we 3D print such a metamaterial with an elastoplastic material (see Supplementary Material). Under quasistatic compression, the bottom floppy mode buckles first, followed by a negative stiffness, until all the diamonds in the first buckled layer reach self-contact (Fig.~\ref{fig:sequential_YB}c top). At this point, the metamaterial stiffens until the critical buckling load of the single floppy mode is reached again (Fig.~\ref{fig:sequential_YB}c bottom). The sequential deformation of the two floppy modes allows going past the isolated inner objects and the curved boundary, and enables stable energy absorption (Fig.~\ref{fig:sequential_YB}d). Such sequential yield buckling of the identical floppy modes can also be designed in a large-scale metamaterial with a higher number of buckling sequences and can be generalized to other elastoplastic materials (see Appendix). However, since the geometry of each floppy mode is identical, the buckling order of the modes is determined by imperfections and boundary conditions.}

\rs{We further demonstrate that the length of floppy modes can harness the buckling order in metamaterials. We design three roughly horizontal floppy chains of different lengths and separate them by rigid layers (Fig.~\ref{fig:sequential_YB}e).} Under quasistatic compression, the straight mode of \rs{shortest length} buckles first, followed by a negative stiffness, until all the diamonds in the first buckled layer reach self-contact (Fig.~\ref{fig:sequential_YB}f top). At this point, the metamaterial stiffens until the critical buckling load of the second-length mode is reached (Fig.~\ref{fig:sequential_YB}f middle). Such sequential buckling repeats until the last mode with the longest length buckles and reaches contact (Fig.~\ref{fig:sequential_YB}f bottom). Due to this sequential buckling, the force curve exhibits a wiggly increasing plateau with tunable local maxima (Fig.~\ref{fig:sequential_YB}g). \rs{In contrast, in a metamaterial of the same geometry, but printed from an elastic material, the three floppy modes buckle almost simultaneously because of their close buckling loads and the positive stiffness at the onset of elastic buckling (see Appendix).} Thus, our designed floppy modes \rs{together with plasticity bring a new approach to design the way the metamaterial collapses under compression and to tailor its nonlinear force displacement curve. Our design tool could complement optimization tools, which so far have relied on a limited set of metamaterial topologies~\cite{bordiga2024automated, meeussen2025textile}.} 

\emph{Matrix-Vector Multiplication --} 
\rs{Beyond the design of nonlinear responses, our approach can also be used to compute using mechanics;} Computing in materia is an emerging direction~\cite{momeni2024trainingphysicalneuralnetworks, yasuda2021mechanical}, studied in a myriad of physical platforms, such as cross-bar arrays~\cite{xia2019memristive}, spintronic devices~\cite{ahn20202d}, electromagnetic~\cite{estep2014magnetic, silva2014performing} and phononic coupled resonators~\cite{okamoto2013coherent}, and microfluidics~\cite{ahrar2023pneumatic}. For these, the ability to perform algebraic operations is paramount, and therefore a crucial challenge is how to achieve matrix manipulations. Recent work has shown that floppy modes can be used to achieve matrix-vector multiplication in mechanical systems~\cite{louvet_matrix_product}, \rs{which could be leveraged to create energy-efficient devices that perform complex tasks, such as battery-less speech detectors~\cite{dubvcek2024sensor} or soft robots with embodied intelligence~\cite{schomaker2024robust}. Yet, design approaches for mechanical matrix-vector products remain largely unexplored.} Here, we demonstrate that \rs{floppy modes and frustrated loops} are an efficient way to \rs{design structures that} perform algebraic operations with multiple inputs and outputs.

\begin{figure}[t!]
\centering
\includegraphics[width=\columnwidth]{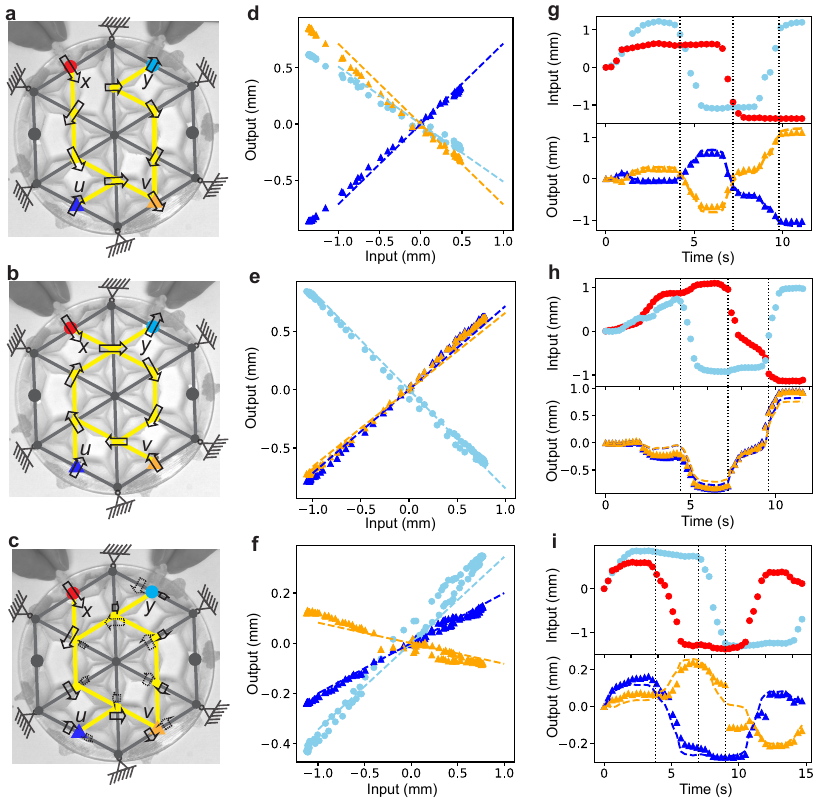}
\caption{\textbf{Matrix-vector multiplication.} \textbf{a-c,} \rs{Experimental setup of open chain (a), even loop (b), and odd loop (c), with the six corner nodes fixed in place. Arrows indicate the decay in displacement along the chain or loop from a single input $x$. Solid and dashed arrows for the odd loop indicate the decay along the two paths. \textbf{d-f} Experimental results (symbols) and theoretical predictions (dashed lines) of output under a single input $x$ with $\alpha=0.92$ for the open chain and even loop, and $\alpha=0.8$ for the odd loop. Colors and markers correspond to the outputs marked in panels a-c. \textbf{g-i,} Experimental measurements (symbols) and theoretical predictions (dashed lines) under simultaneous actuation of the two inputs $x$ and $y$.}}
\label{fig:Matrix_manip}
\end{figure}

\rs{We focus on one of the most basic matrix-vector multiplication: the multiplication of a $2\times 1$ vector by a $2\times 2$ matrix, $\begin{pmatrix} u & v \end{pmatrix}^T = M \begin{pmatrix} x & y \end{pmatrix}^T$. To do this, we construct a family of minimal metamaterials made from six triangular blocks (Fig.~\ref{fig:Matrix_manip}a-c). They all include either an open chain or a loop}. We fix the hexagon's six corner nodes and locally actuate the \rs{top-left and top-right edge nodes, as input displacements $x$ and $y$. We isolate the left and right edge nodes with two $T_1$ blocks to have only two outputs, $u$ and $v$, at the bottom-left and bottom-right edges. The matrix $M$ hence encodes the deformation of the chain or loop.} 

\rs{The ideal models of rigid bonds, realized above using LEGO\textsuperscript{\tiny\textregistered}, are either purely floppy---for open chains and even loops---or strictly rigid---for odd loops. Hence, the entries of $M$ would be limited to either $\pm 1$ or identically zero, respectively. Luckily however, 3D-printed metamaterials are made of hinges that shear and extend~\cite{coulais2016periodic,Coulais_NatPhys2018, meeussen2020supertriangles, czajkowski_conformal_2022}. These elastic distortions induce a decay in displacement by some factor $\alpha < 1$ from one node to the next and, in turn, non-trivial output displacements (Fig.~\ref{fig:Matrix_manip}a-c and Video~4). These displacements add up to form the matrix $M$ with coefficients that are rational functions of $\alpha$ which are set by the length and the topology of the loop; One can determine the sign of the interaction by patterning with an odd number of antiferromagnetic interactions---which flips the displacement's direction. One can determine the value of the coefficient by choosing the length of the chain (see Appendix). Hence, the combinatorial design of the loops yields large design freedom for the matrix $M$.}

\rs{We conclude by testing the matrix-vector product multiplication experimentally. We actuate each of the inputs $x$ and $y$ separately and observe linear behavior of each of the outputs $u$ and $v$, as expected (Fig.~\ref{fig:Matrix_manip}d-f). Remarkably, our measurements are well captured by our analytical predictions, both in the case of single (Fig.~\ref{fig:Matrix_manip}d-f), or combined input displacements (Fig.~\ref{fig:Matrix_manip}g-i). Our approach can readily be generalized to larger matrices (see Supplementary Material) and further design freedom can be gained by considering 3D stacks of combinatorial metamaterials, each made of multiple minimal hexagons (see Supplementary Material). Thus, our combinatorial strategy is a powerful way to design on demand matrix-vector multiplication.}

We introduced a combinatorial approach using unimodal and bimodal triangular building blocks to create metamaterials with arbitrary shapes of multiple floppy modes and frustrated loops. We demonstrated that curved floppy modes together with yield buckling can be used to achieve sequential deformations, and that frustrated loops can be exploited to achieve matrix-vector multiplication. Our findings enrich the toolbox of sequential metamaterial design and lead to various applications in shock and vibration damping, shape changing, and mechanical computing. \rs{The geometric framework we have introduced opens exciting questions: how to leverage geometric nonlinearities to create even more advanced functions such as non-reciprocal computing~\cite{coulais2017static} and memory devices~\cite{guo2023non}? How to harness material nonlinearities such as viscoelasticity, plasticity, and fatigue to endow mechanical computing with memory~\cite{du2025metamaterialslearnchangeshape, comoretto2025embodying}?}

\emph{Acknowledgments --} 
We thank Chaviva Sirote-Katz, Daniela Kraft, Dor Shohat, Julio Melio, Marc Serra-Garcia, and Yuan Zhou for inspiring discussions and Bat-El Pinchasik and Yaara Shokef for technical assistance. T.A.S. and Y.S. thank the Institute of Physics at the University of Amsterdam for its hospitality. T.A.S. acknowledges funding from the Israeli Ministry of Energy and Infrastructure's scholarship program for undergraduate students in the field of energy. W.L. and C.C. acknowledge funding from the European Research Council under grant agreement 852587 and the Netherlands Organisation for Scientific Research under grant agreement NWO TTW 17883. C.C. acknowledges funding from the Netherlands Organisation for Scientific Research under grant agreement NWO under grant agreement VI.Vidi 2131313. This research was supported in part by the Israel Science Foundation Grant No. 1899/20. 

\emph{Code and Data Availability --} The supplementary videos as well as all the data and computer codes supporting this work are available at \url{https://zenodo.org/records/17233065}.

\clearpage

\section{Appendix}

\subsection{Bounds on Number of Floppy Modes} \label{app: spin bounds}

To maximize the number $F$ of floppy modes, we minimize the number $R$ of rigid chains and maximize the number $L$ of loops. The minimal value of $R$ is zero. We now derive two upper bounds on $L$. The first stems from the fact that each loop surrounds at least one internal corner node, thus $L \le \frac{N+2-P}{2}$, which leads to $F \le N_1 + 1$.

For the second bound, we start with a lattice of $N_2=0$ and all triangles as $T_1$; In this limit, the maximal number of loops gives $L \le \frac{N}{6}$, by arranging the loops in a honeycomb formation. Now, as we gradually switch $T_1$ blocks to $T_2$, the most efficient way to create new loops is that two loops share each $T_2$ block. Each new loop needs six bonds, thus $L \le \frac{N}{6} + \frac{N_2}{3}$, and $F \le \frac{2N_1}{3} + \frac{P}{2}$. The crossover between these two bounds occurs at $N_1 = \frac{3P}{2}-3$, and we obtain the upper bound given in Eq.~(\ref{eq:bounds}).

For given lattice size and shape, the system is more rigid as there are more $T_2$ than $T_1$ blocks. We obtain the minimal number $N_2$ of $T_2$ blocks or maximal number $N_1$ of $T_1$ blocks required for rigidity by requiring that all $E$ edge nodes belong to a single chain with at least one odd loop. Thus, the number of internal bonds must be at least $E$, with $E-1$ bonds connecting all nodes and one bond closing an odd loop. Thus, mechanical rigidity can be achieved when $N_1 \le \frac{N-P}{2}$. In this case $F=0$. As we increase $N_1$, we remove internal bonds. There are no loops, and this removal of bonds does not create loops. Thus, by substituting in the expression for the number of modes, we obtain the lower bound given in Eq.~(\ref{eq:bounds}).

\begin{figure}[t]
\centering
\includegraphics[width=\columnwidth]{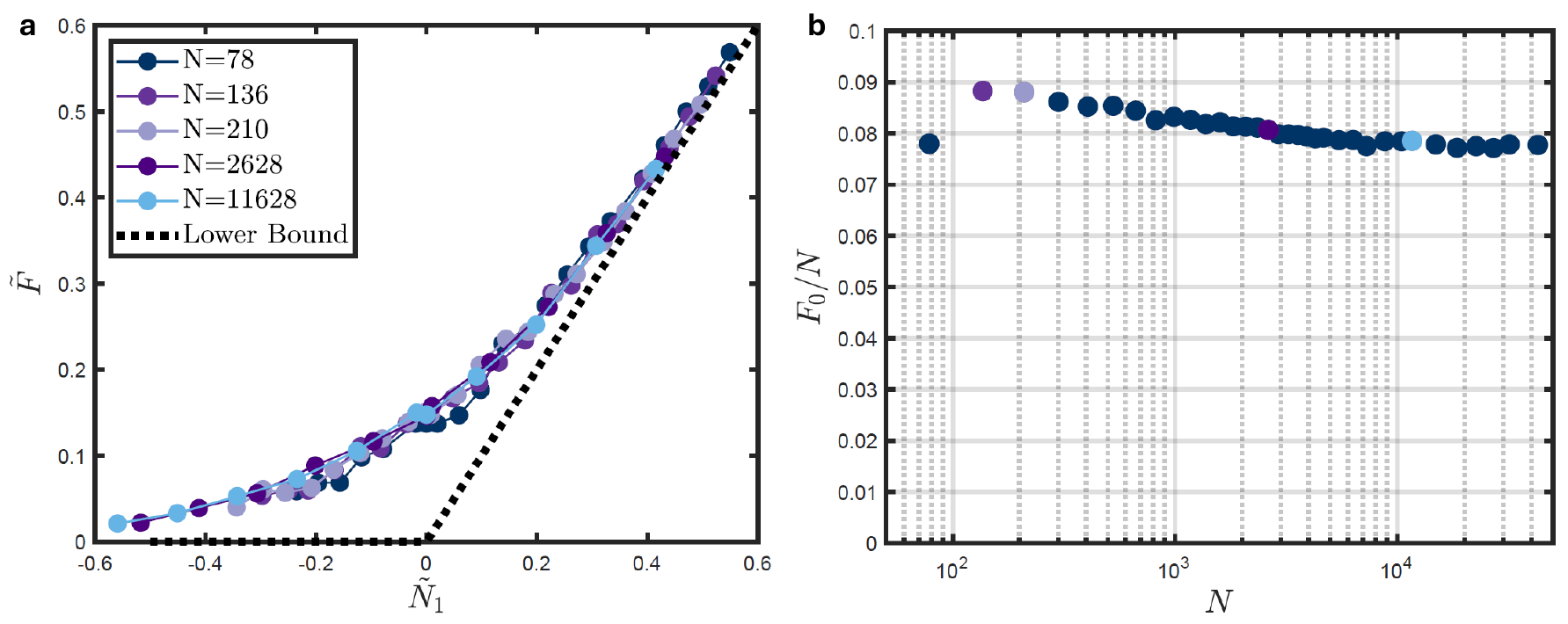}
\caption{\textbf{Average number of floppy modes in random lattices.} \textbf{a,} Results for different system sizes normalized as $\tilde{F} = F / [(N+P)/2]$ and $\tilde{N_1} = [N_1 - (N-P)/2] / [(N+P)/2]$, so that the lower bounds coincide. \textbf{b,} Normalized number of floppy modes at the onset of rigidity, $\tilde{N}_1=0$.}
\label{fig:statistical-results}
\end{figure}

In the large system size limit, $N \gg 1$, since $P \ll N$, the lower bound is approximated by
\begin{equation}
F \ge \begin{cases}
0, & N_1 \le \frac{N}{2} \\
N_1 - \frac{N}{2} , & N_1 \ge \frac{N}{2} ,
\end{cases}
\end{equation}
and the upper bound by $F \le \frac{2N_1}{3}$. As system size increases, the average number of floppy modes in randomly generated lattices collapses to a single curve (Fig.~\ref{fig:statistical-results}).

\rs{For high fractions $N_1/N$ of $T_1$ blocks, random lattices are not likely to possess loops, thus $L \ll N$, and clearly also $L - R \ll N$. Thus, the typical number of floppy modes is close to the lower bound, as seen in Fig.~\ref{fig:1}d.}

\rs{\subsection{Number of Floppy Modes in Layered Systems}}

\rs{For a metamaterial made of $K$ coupled layers, each with $N$ triangles and perimeter $P$, the total number of edge nodes is $\left( \frac{3N}{2} + \frac{P}{2} \right) K$, and the total number of bonds between them is $N_1+2N_2+C$, where $C$ is the number of inter-layer connectors. Note that $N_1$ and $N_2$ are the total numbers of $T_1$ and $T_2$ triangles across all layers. Thus, their sum is given by $N_1 + N_2 = N K$. Following the same counting arguments presented above for the single-layer case, we now obtain $F = N_1 - \frac{NK}{2} + \frac{PK}{2} - C + L -R$, where $L$ and $R$ count loops and rigid chains that may now also close by crossing between multiple layers.}

\rs{\subsection{Matrix-Vector Multiplication}}

\rs{For a single input $x$ or $y$, we denote the linear response between the input and outputs as $\begin{pmatrix} u & v \end{pmatrix}^T=A\begin{pmatrix} x & y \end{pmatrix}^T$ and $\begin{pmatrix} x_0 & y_0 \end{pmatrix}^T=H\begin{pmatrix} x & y \end{pmatrix}^T$, where $x_0$ and $y_0$ are the displacements generated at the other output node.} 

\rs{To find the matrix $M$ relating all input and output displacements $\vec{U}=\begin{pmatrix} x & y & u & v \end{pmatrix}^T$, we consider the reaction forces at the inputs and outputs, $\vec{F}=\begin{pmatrix} f_x & f_y & f_u & f_v \end{pmatrix}^T$. The compliance matrix $\mathcal{C}$ relates them by $\vec{U} = \mathcal{C} \vec{F}$.}

\rs{For a single input $x$, $\vec{F} = \begin{pmatrix} f_x & 0 & 0 & 0 \end{pmatrix}^T$, thus $\vec{U} =\begin{pmatrix} \mathcal{C}_{xx} & \mathcal{C}_{yx} & \mathcal{C}_{ux} & \mathcal{C}_{vx} \end{pmatrix}^T f_x$. However, in this case we may also write $\vec{U} = \begin{pmatrix} 1 & H_{yx} & A_{ux} & A_{vx} \end{pmatrix}^T x$. Therefore, $\vec{\mathcal{C}}_x \equiv \begin{pmatrix} \mathcal{C}_{xx} & \mathcal{C}_{yx} & \mathcal{C}_{ux} & \mathcal{C}_{vx} \end{pmatrix}^T = \begin{pmatrix} 1 & H_{yx} & A_{ux} & A_{vx} \end{pmatrix}^T \mathcal{C}_{xx}$. For a single input $y$, we similarly obtain \\ $\vec{\mathcal{C}}_y \equiv \begin{pmatrix} \mathcal{C}_{xy} & \mathcal{C}_{yy} & \mathcal{C}_{uy} & \mathcal{C}_{vy} \end{pmatrix}^T = \begin{pmatrix} H_{xy} & 1 & A_{uy} & A_{vy} \end{pmatrix}^T \mathcal{C}_{yy}$.}

\rs{For two inputs $x$ and $y$, $\vec{F} = \begin{pmatrix} f_x & f_y & 0 & 0 \end{pmatrix}^T$, and $\vec{U} = \begin{pmatrix} \vec{\mathcal{C}}_x & \vec{\mathcal{C}}_y \end{pmatrix} \begin{pmatrix} f_x & f_y \end{pmatrix}^T$. Namely, \\ $\begin{pmatrix} x \\ y \end{pmatrix} = \begin{pmatrix} \mathcal{C}_{xx} & \mathcal{C}_{xy} \\ \mathcal{C}_{yx} & \mathcal{C}_{yy} \end{pmatrix} \begin{pmatrix} f_x \\ f_y \end{pmatrix}$ and $\begin{pmatrix} u \\ v \end{pmatrix} = \begin{pmatrix} \mathcal{C}_{ux} & \mathcal{C}_{uy} \\ \mathcal{C}_{vx} & \mathcal{C}_{vy} \end{pmatrix} \begin{pmatrix} f_x \\ f_y \end{pmatrix}$. We rewrite these two relations as, \\ $\begin{pmatrix} u \\ v \end{pmatrix} = \begin{pmatrix} \mathcal{C}_{ux} & \mathcal{C}_{uy} \\ \mathcal{C}_{vx} & \mathcal{C}_{vy} \end{pmatrix} \begin{pmatrix} \mathcal{C}_{xx} & \mathcal{C}_{xy} \\ \mathcal{C}_{yx} & \mathcal{C}_{yy} \end{pmatrix}^{-1} \begin{pmatrix} x \\ y \end{pmatrix}=M\begin{pmatrix} x \\ y \end{pmatrix}$. Now, \scalebox{0.75}{$\begin{pmatrix} \mathcal{C}_{ux} & \mathcal{C}_{uy} \\ \mathcal{C}_{vx} & \mathcal{C}_{vy} \end{pmatrix} \begin{pmatrix} \mathcal{C}_{xx} & \mathcal{C}_{xy} \\ \mathcal{C}_{yx} & \mathcal{C}_{yy} \end{pmatrix}^{-1} = \begin{pmatrix} A_{ux} & A_{uy} \\ A_{vx} & A_{vy} \end{pmatrix} \begin{pmatrix} \mathcal{C}_{xx} \\ \mathcal{C}_{yy} \end{pmatrix} \begin{pmatrix} 1 & H_{xy} \\ H_{yx} & 1 \end{pmatrix}^{-1} \begin{pmatrix} \mathcal{C}_{xx} \\ \mathcal{C}_{yy} \end{pmatrix}^{-1}$}. Therefore, $M = A H^{-1}$.}

\rs{The same method can be generalized to any number of inputs $\vec{X} = \begin{pmatrix} x_1 &x_2 &...& x_n \end{pmatrix}$ and outputs $\vec{O}=\begin{pmatrix} u_1 & u_2 &...& u_m \end{pmatrix}$, leading to $\vec{O} = A_{m\times n} H_{n\times n}^{-1} \vec{X}$.}

\rs{\subsection{Matrices Design Space}}

\rs{To theoretically calculate the matrices $A$ and $H$ based on the metamaterial's design, we assume the displacement decays by a factor $\alpha$ in each bar of the metamaterial. The factor $\alpha$ is determined by the geometry of the metamaterials, which is related to the hinge deformation. Due to frustration, we expect a smaller factor $\alpha$ in the odd loop than in the open chain and even loop. In each triangular unit, if the path from input to output includes one bar, the direction of displacement is conserved, while if there are two bars, it is flipped. Therefore, for an open chain, $A_{ij}$ and $H_{ij}$ are both given by $(-1)^{t_{ij}+b_{ij}+1} \alpha^{b_{ij}}$, where $t_{ij}$ is the number of triangles along the path between input $i$ and output $j$, and $b_{ij}$ is the number of bonds along this path. For hexagons such as those given in Fig.~\ref{fig:Matrix_manip}, the numbers of triangles are $t_{xy}=2$, $t_{xu}=t_{yv}=3$, $t_{xv}=t_{yu}=4$.} 

\rs{For hexagons with all possible open chains connecting $x$ and $y$, the numbers of bonds may take the values $b_{xy}=6, 7, 8, 9$, $b_{xu}, b_{yv}=3, 4, 5$, and $b_{xv}, b_{yu}=4, 5, 6, 7$. For the specific design given in Fig.~\ref{fig:Matrix_manip}a, this leads to, \\
$A_{open}=\begin{pmatrix}
\alpha^4 & -\alpha^6 \\
-\alpha^4 & \alpha^4
\end{pmatrix}$, $H_{open}=\begin{pmatrix}
1 & -\alpha^8 \\
-\alpha^8 & 1
\end{pmatrix}$.}

\rs{For a metamaterial with a loop, there are two paths between the input and output. If the loop is even, the two paths lead to output displacements in the same direction, and we assume that the magnitude of this displacement is dominated by the shorter of the two paths. For hexagons with even loops, as defined in the Main Text, the possible numbers of bonds along the shorter path are, $b_{xy}=2, 3, 4$, $b_{xu}, b_{yv}=3, 4, 5$, and $b_{xv}, b_{yu}=4, 5, 6$, and for the design given in Fig.~\ref{fig:Matrix_manip}b, we have,
$A_{even}=\begin{pmatrix}
\alpha^4 & -\alpha^4 \\
\alpha^5 & -\alpha^5
\end{pmatrix}$, $H_{even}=\begin{pmatrix}
1 & -\alpha^2 \\
-\alpha^2 & 1
\end{pmatrix}$.}

\rs{In a metamaterial with an odd loop, the two paths lead to output displacements in opposite directions. We assume that these add up such that $A_{ij}$ and $H_{ij}$ are given by the sum of $(-1)^{t_{ij}+b_{ij}+1} \alpha^{b_{ij}}$ with $t_{ij}$ and $b_{ij}$ of the two paths. The design space of a hexagon with an odd loop has, $b_{xy} = (2,7), (2,9), (3,6), (3,8), (4,7)$, $b_{xu}, b_{yv}=(3,6), (3,8), (4,7), (5,6)$, and $b_{xv}, b_{yu} = (4,5), (4,7), (5,6)$, where pairs indicate the values of $b_{ij}$ along the two paths.  Tracking the two paths from each input to each output node for the structure of Fig.~\ref{fig:Matrix_manip}c, we obtain $A_{odd}=\begin{pmatrix}
  \alpha^4-\alpha^7 &\alpha^5-\alpha^6\\
-\alpha^4+\alpha^5 & \alpha^4-\alpha^5
\end{pmatrix}$, \\ 
$H_{odd}=\begin{pmatrix}
1 & \alpha^3-\alpha^8 \\
\alpha^3-\alpha^8 & 1
\end{pmatrix}$.}

\rs{See Supplementary Material for realizations of all possible values of $b_{ij}$ listed above.}

\vspace{0.5cm}

\rs{\subsection{Sequential Buckling vs. Simultaneous Buckling}}

\begin{figure}[htb]
\centering
\includegraphics[width=\columnwidth]{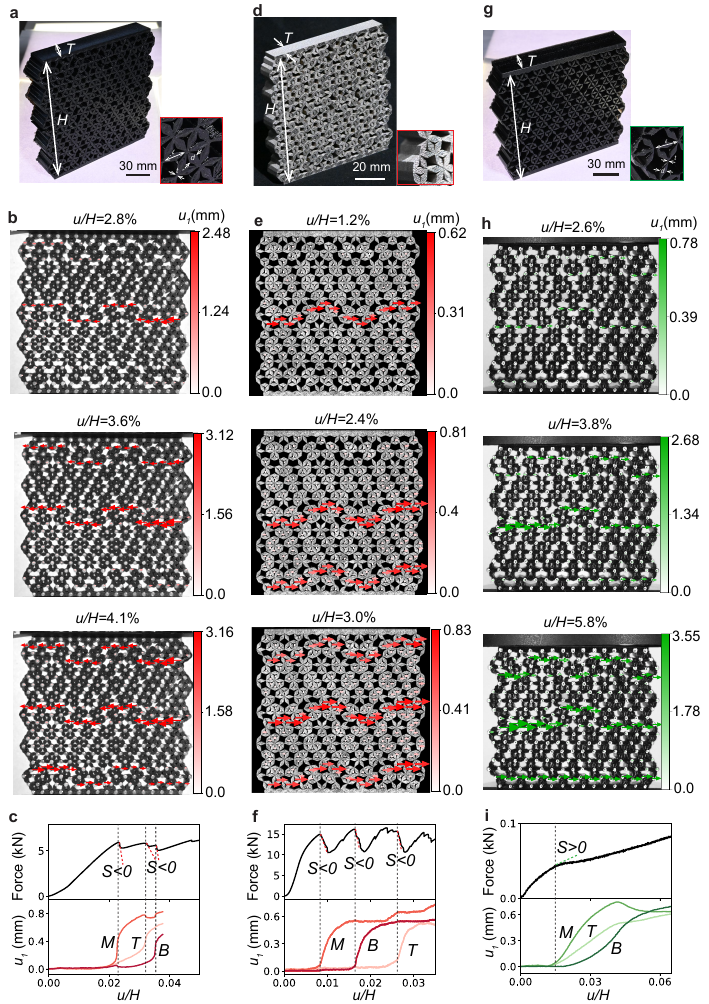}
\caption{\rs{\textbf{Sequential actuation of floppy modes.} \textbf{a,} Metamaterial with three identical floppy modes printed of plastic Nylon. \textbf{b,}  Compression of the metamaterial at different strokes. Color-coded arrows in b, e, and h indicate horizontal displacement. \textbf{c,} Corresponding vertical force and average horizontal displacement, $u_1$ of each floppy chain vs. global compressing stroke $u/H$. \textbf{d,} Metamaterial with the same design as panel a, printed of stainless steel. \textbf{e and f,} Corresponding compression at three different strokes and force-displacement curves. \textbf{g,} 3D printed elastic metamaterial with three floppy chains of varying length. \textbf{h and i,} Elastic metamaterial at different stages of compression and force-displacement curves. $B$, $M$, and $T$ in c, f, and i denote the bottom, middle, and top floppy chains. The sign of the slope ($S$) of force vs. compression is detected (dashed lines).} }
\label{fig:sequential-snakes}
\end{figure}

\rs{We show that sequential yield buckling with arbitrarily many orders can be generalized in any other elastoplastic materials with periodically arranged floppy modes (Fig.~\ref{fig:sequential-snakes}a-f), whereas in elastic metamaterials, floppy modes buckle simultaneously, regardless of length differences (Fig.~\ref{fig:sequential-snakes}g-i).}

\clearpage

\bibliography{bibliography}

\end{document}